\documentstyle[12pt,fullpage]{article}
\def\be{\begin{equation}}
\def\ee{\end{equation}}
\def\beq{\begin{eqnarray}}
\def\eeq{\end{eqnarray}}

\begin{document}
\begin{flushright}
TIFR/TH/97-24
\end{flushright}
\bigskip
\begin{center}
{\Large{\bf New Physics from HERA?}} \\[1cm]
D.P. Roy \\
Tata Institute of Fundamental Research \\
Homi Bhabha Road, Colaba \\
Mumbai - 400 005, INDIA \\[2cm]
\underbar{\bf Abstract}
\end{center}
\bigskip

The H1 and ZEUS experiments at the HERA $e^\pm p$ collider at Hamburg
have recently reported some anomalous hard-scattering events, which
could be 
indicative of new physics beyond the standard model.  I have tried to
discuss in a nonspecialist language the significance of this result
for particle physics along with its interpretation in terms of the
currently popular extensions of the standard model.

\newpage

\noindent {\bf Basic Constituents of Matter:}  Our understanding of
the basic constituents of matter has undergone two revolutionary
changes during this century.  The first was the Rutherford scattering
experiment of 1911, which showed that the atom is made up of a compact
nucleus (containing protons and neutrons), surrounded by the tiny
electrons.  The second was the electron scattering experiment of 1968
at Stanford, performed at a thousand times higher energy, which showed
that the proton and neutron are themselves made up of tiny
constituents called quarks.  Both proton and neutron are composed of
two types of quarks called up and down.  Thus these quarks along with
the electrons constitute all the visible matter of the Universe.

The electron along with its massless and neutral partner, neutrino,
are called leptons.  There are two heavier pairs of leptons as well as
quarks.  But they decay promptly into the light ones and hence do not
occur freely in nature.  All these quarks and leptons are fermions,
since they carry spin 1/2 in natural units,
\be
\hbar = c = 1,
\ee
where $\hbar$ and $c$ denote Plank's constant and the velocity of
light. 
\bigskip

\noindent {\bf Basic Interactions:} Apart from gravity, whose
influence in the subatomic world is negligible, there are 3 basic
interactions -- strong, electromagnetic and weak.  They are all gauge
interactions, mediated by spin 1 particles called gauge bosons.  The
quarks have strong interaction, mediated by gluons, which is
responsible for binding them together inside proton and neutron.  This
is analogous to the electromagnetic interaction between the quarks and
the electrons, mediated by the photon, which binds them together in the
atom.  All the quarks and leptons including the neutrinos experience
weak interaction, which is responsible for nuclear decay.  Unlike the
strong and the electromagnetic interactions, which are mediated by
massless gauge bosons, the weak interaction is mediated by massive
gauge bosons called $W$ and $Z$.  The theory of these basic
constituents of matter along with their gauge interactions is known as
the Standard Model (SM).
\bigskip

\noindent {\bf GeV to TeV Energies:}  It follows from the Uncertainty
Principle that the smaller the distance to be probed, the larger must
be the beam energy.  Thus to probe inside a proton of dimension about
1 fm $(10^{-13} \ {\rm cm})$, one needs an electron beam energy
\be
E_e > \hbar c/1 \ {\rm fm} \ {\rm i.e.} \ E_e > 1 \ {\rm GeV},
\ee
where a GeV (Gega electron Volt) is the energy acquired by the
electron after passing through $10^9$ Volts.  It is customary to use
the natural system of units (1), in which case the GeV becomes a
convenient unit for mass, energy and momentum.  The mass of the proton
is about 1 GeV.

It is the multi-GeV electron beam energy, that enabled the above
mentioned Stanford experiment of 1968 to probe the structure of the
proton.  Thanks to the colliding beam technology, we have seen a
thousand fold increase of the invariant energy, from the GeV to the
TeV scale, since then.  The invariant energy corresponds to the energy
measured in the centre of momentum (CM) frame, which is the effective
energy available for particle creation.  This has led to a string of
discoveries over the past 25 years.  The charm and bottom quarks, the
tau lepton and the gluon were discovered during the seventies, thanks
mainly to the electron-positron colliders at Stanford and Hamburg.
This was followed by the discovery of the massive $W,Z$ bosons with
masses 
\be
M_W \simeq 80 \ {\rm GeV}, \ \ \ M_Z \simeq 91 \ {\rm GeV},
\ee
at the CERN proton-antiproton collider in 1983.  Finally the last and
the heaviest member of the quark family, the top quark with mass 
\be
M_t \simeq 175 \ {\rm GeV},
\ee
was discovered at the Tevatron proton-antiproton collider at Fermilab
in 1995.  Thus we have seen all the basic constituents of matter by
now along with the carriers of the basic interactions.  Moreover the
large electron-positron (LEP) collider at CERN has made it possible to
check the predictions of the standard model, including quantum corrections, to great accuracy.
In particular the measured masses and widths of the $W$ and $Z$ bosons
are in remarkable agreement with the predictions of the unified
electroweak theory.
\bigskip

\noindent {\bf What Next?:}  The story does not end here, however.  A
consistent theory of the massive gauge bosons, $W$ and $Z$, requires
the presence of one or more scalar (spin 0) particles of comparable
mass.  These are called Higgs bosons.  But the story does not end here
either.  In the absence of a protecting symmetry, the scalar masses
are driven to infinity by the quantum corrections!  Thus to control these
scalar particle masses one invokes supersymmetry (SUSY) -- a symmetry
between fermions and bosons.  This implies the existence of scalar
superpartners of quarks and leptons as well as fermionic partners of
the gauge and Higgs bosons, again in the mass range of $W$ and $Z$
bosons -- i.e. around $10^2$ GeV (3).  Thus the Higgs and the SUSY
particles represent a minimal set of missing pieces, required for a
consistent theory of fundamental particles.  Besides the lightest SUSY
particles (LSP) is a promising candidate for the dark matter of the
Universe.  Thus the immediate goal of particle physics is largely
focussed on these particles.  On the other hand the long term
objectives are the unification of strong along with the electroweak
interaction in the form of a Grand Unified Theory (GUT) and
unltimately to rope in gravity as well.  But it is fair to state here
that there is as yet no experimental evidence for Higgs or SUSY
particles, or for that matter any other form of new physics beyond the
standard model. 
\bigskip

\noindent {\bf The HERA $e^\pm p$ Collider:}  The latest colliding
machine is HERA at Hamburg, operating since 1993, where a beam of
electron $e^-$ (or positron $e^+$) collides head on against a beam of
proton.  Most of the data so far has been taken with the $e^+$ beam.
The beam energies are 28 and 820 GeV for the $e^\pm$ and proton,
compared to which the corresponding particle masses are negligible.
Thus the CM energy is
\be
\sqrt{s} = 2\sqrt{E_e E_p} = 2\sqrt{28 \times 820} \simeq 300 \ {\rm
GeV}. 
\ee
There are two detectors engaged in recording $e^+ p$ collision events
at HERA, named ZEUS and H1.  HERA and ZEUS are named after the famous 
Greek deities (Hypertext Webster Gateway defines HERA as the sister
and wife of ZEUS!), while H1 has evidently a more mundane origin.

The higher machine energy has made it possible to probe the quark and
gluon distributions inside the proton much more precisely at HERA than
in earlier experiments.  But more importantly, a recent analysis of
the data collected by these two experiments during 1994-96 have shown
about 10 anomalous events, which could be suggestive of New Physics
beyond the SM.  The results were presented in a joint seminar by the
two groups at Hamburg followed by a press report last February,
which have recently been published [1,2].  While the event sample is
still very small, it has generated a good deal of excitement around
the world along with a flurry of e-prints, of which only a partial
list is given in refs. [3-8].  In order to discuss the significance of
these events and their theoretical interpretation, it will help to
briefly summarise the kinematics of $ep$ scattering at HERA.

Fig. 1 shows a space-time picture of $ep$ scattering, with time axis
running vertically upwards.  The positron interacts with a quark
carrying a 4-momentum fraction $x$ of the proton.  Thus the invariant
energy of the $eq$ pair is
\be
M = \sqrt{s \cdot x}.
\ee
Over the hard scattering region of interest the measured quark
momentum distribution inside the proton roughly corresponds to 
\be
\langle x\rangle \simeq 0.1, \ \ \ \langle M\rangle \simeq 100 \ {\rm
GeV}. 
\ee
The squared 4-momentum transfer between the incident and outgoing
positron (or quark) is denoted by $Q^2$.  This is related to the CM
scattering angle $\theta^\star$ of the $e-q$ pair via
\be
Q^2 = y M^2, \ \ \ y = (1 - \cos\theta^\star)/2,
\ee
i.e.
\be
0 < Q^2 < M^2.
\ee

The SM interaction between the $e^\pm$ and the quark is the
electroweak interaction mediated by the photon $(\gamma)$ and the
massive $Z$ boson exchanges as shown in Fig. 1(a).  The $Q^2$
dependence resulting from the $\gamma$ and $Z$ propagators are
\be
{d\sigma_\gamma (M) \over dQ^2} \propto {1 \over Q^4}, \ \ \
{d\sigma_Z (M) \over dQ^2} \propto {1 \over (Q^2 + M^2_Z)^2}.
\ee
The corresponding $M$-integrated scattering cross-sections will fall
even faster with increasing $Q^2$ because of the kinematic
constraint (9).

In contrast the presence of a heavy leptoquark state (a hypothetical
particle coupling to lepton and quark), illustrated in
Fig. 1(b), would signal events which are clustered around a high
invariant mass of the $e-q$ pair
\be
M \simeq M_{\ell q}, \ \ \ x \simeq M^2_{\ell q}/s.
\ee
Moreover they would have a flat $Q^2$ distribution as the sense of
the original direction is lost after the formation of the leptoquark.
In particular a scalar leptoquark would have an isotropic decay and
hence a flat $Q^2$ distribution via (8).

Each of the two HERA experiments shows an excess of high $Q^2$
events over the SM prediction, indicating an anomalous hard component in $e^+ p$ scattering.
\bigskip

\noindent {\bf High $Q^2$ Events:}  The ZEUS experiment shows 5
events against the SM prediction of 2 for $Q^2 > 20000 \ {\rm
GeV}^2$.  Moreover the excess of 3 events are consistent with a
common $e^+ q$ invariant mass $M \simeq 200 \ {\rm GeV}$.  One of
these events is shown in Fig. 2.  The inset on top clearly shows the
scattered positron and quark-jet in the scattering plane, while the
remaining proton fragments escape in the beam pipe.  The bottom inset
shows their back-to-back configuration in the transverse plane, as
required for transverse momentum balance.  The magnitude of their
transverse momenta are shown in the lego plot as $E_T$, i.e.
\be
E^e_T \simeq E^q_T \simeq 100 \ {\rm GeV}.
\ee
In fact from this figure one can easily reconstruct the rough
magnitudes of $M$ and $Q^2$.  It has evidently an unlikely
kinematic configuration for SM scattering as it corresponds to a very
hard quark $(x \simeq 0.5)$ and a backward $e^+ q$ scattering in the
CM frame $(\theta^\star > 90^\circ)$.

The H1 experiment shows 12 events against a SM prediction of 5 for
$Q^2 > 15000 \ {\rm GeV}^2$, as shown in Fig. 3.  Moreover the
excess of 7 events are consistent with a common invariant mass $M =
200 \pm 20 \ {\rm GeV}$, as indicated by arrows in this figure.
Indeed the common mass and the flat distribution over a very wide
range of $Q^2$ are suggestive of isotropic decay of a leptoquark state
as discussed above.
\bigskip

\noindent {\bf Contact Interaction:}  Apart from leptoquark
production, there is another mechanism suggesting a flatter $Q^2$
dependence than the SM.  This corresponds to the exchange of a very
massive particle between the positron and quark in Fig. 1(a) -- e.g. a
heavy gauge boson $Z'$ occurring in many extensions of the SM.  This
is called contact interaction, since the exchange of a heavy particle
is restricted to a tiny range $(\hbar/M_{Z'} c)$ via the Uncertainty
Principle.  Clearly the resulting cross-section of eq. (10) will be
flat in $Q^2$ for $M^2_{Z'} \gg Q^2$.  However, this
interpretation is disfavoured on three counts.
(i) The size of the effect required to explain the HERA
events seems to be larger than the upper limit placed on this quantity
from LEP and Tevatron collider data.  (ii) It favours the standard $M$
distribution, as suggested by the measured quark distribution inside
the proton (7), in stead of a clustering of events at a high value of
$M$.  (iii) For the above reason the $M$ integrated cross-section
falls significantly with increasing $Q^2$.  Consequently this
interpretation of the anomalous HERA events is strongly disfavoured
[4,7], although it may not be completely ruled out [8].
\bigskip

\noindent {\bf Leptoquarks:}  The kinematic distribution of the
anomalous HERA events clearly favours the formation and decay of a
bound state in the $e^+ q$ system -- i.e. a generic leptoquark [3-6].
So it is natural to ask whether the various extensions of the SM
discussed above can naturally accommodate such a leptoquark.  The
leptons and quarks are unified in GUT, which naturally predict
leptoquark states both as gauge bosons and Higgs scalars.  However the
exchange of these objects in the GUT generally leads to lepton and
baryon number (or equivalently the quark number) violating interactions,
and in particular to proton decay.  Thus the stability of proton
implies these objects to be very heavy $(M_{\ell q} > 10^{15} \ {\rm
GeV})$, which puts them far beyond the reach of present or foreseeable
future experiments.

A more plausible scenario for such generic leptoquarks is the scalar
superpartner of quark (squark) in the so called $R$-parity violating
SUSY model [3-6].  As mentioned earlier, these particles are expected
to occur in the right mass range of a few hundred GeV.  In general
they can have both lepton and baryon number violating Yukawa couplings
and mediate proton decay.  Usually these couplings are set to be zero
by assuming $R$-parity conservation.  Unlike the gauge couplings,
however, these Yukawa couplings are not connected to any symmetry
consideration.  Therefore one can assume a finite value for the lepton
number violating coupling while setting the baryon number violating
one to zero.   The former ensures squark coupling to the $e^+ q$
channel, while the latter prevents proton decay.  Thus in the
$R$-parity violating SUSY model the squark can masquerade as a
leptoquark and account for the anomalous HERA events.  The price one
has to pay is that in this case the lightest SUSY particle (LSP) will
no longer be stable and hence not a candidate for the dark matter.  It
is equally possible of course that these generic leptoquarks could
have an origin outside the currently popular extensions of the SM.
\bigskip

\noindent {\bf Concluding Remarks:}  The most plausible explanation of
the anomalous HERA events within the SM is the statistical fluctuation
of a small number of events into an unlikely configuration.  Using the
standard statistical meathods, the probability of this fluctuation can
be estimated to be less than 1\% for each experiment [1,2].  This
corresponds to nearly a 3 sigma deviation, which is by no means a
definitive signal for new physics.  What gives credence to this result
is of course its simultaneous observation in two experiments.
Nonetheless one should bear in mind the risks of statistical
fluctuation while dealing with so few signal events.  The ongoing
experiment at HERA is expected to double the data sample by the end of
next year.  Moreover most of the new mechanisms for these events will
imply visible effects in the dilepton channel in the present and
forthcoming Tevatron collider data.  Thus one expects a clear picture
to emerge in a year or two.

I thank Prof. Virendra Singh for asking me to undertake this
exercise and a careful reading of the manuscript. 
\bigskip

\noindent \underbar{\bf References}
\bigskip

\begin{enumerate}
\item[{1.}] H1 Collaboration: Zeit. Phys. C74, 191 (1977).
\item[{2.}] ZEUS Collaboration: Zeit. Phys. C74, 207 (1997).
\item[{3.}] D. Choudhury and S. Raychaudhuri, hep-ph/9702392.
\item[{4.}] G. Altarelli et al., hep-ph/9703276.
\item[{5.}] H. Dreiner and P. Morawitz, hep-ph/9703279.
\item[{6.}] J. Kalinowski et al., hep-ph/9703288.
\item[{7.}] K.S. Babu et al., hep-ph/9703299.
\item[{8.}] V. Barger et al., hep-ph/9703311.
\end{enumerate}

\newpage

\begin{center}
{\bf Figure Captions}
\end{center}
\bigskip

\begin{enumerate}
\item[{Fig. 1.}] Space-time picture of $e^+ p$ scattering (a) via
photon and $Z$ boson exchanges (SM), and (b) via a leptoquark state. 
\item[{Fig. 2.}] One of the anomalous high $Q^2$ events from the
ZEUS experiment [2].  The top (bottom) inset shows the scattered
positron and quark-jet in the scattering (tansverse) plane.  The lego
plot shows the transverse energy distribution in azimuthal angle and
rapidity $(\eta = -\ell n \tan \theta/2)$.
\item[{Fig. 3.}] The $Q^2$ distribution of the H1 events showing an
excess of 7 events over the SM prediction of 5 for $Q^2 > 15000 \
{\rm GeV}^2$.  The arrows indicate the 7 events having a common $M =
200 \pm 20 \ {\rm GeV}$ [1]. 
\end{enumerate}

\end{document}